\documentclass[preprint,tightenlines,nofootinbib,showpacs]{revtex4}
\usepackage{graphicx}

\newcommand{\be}{\begin{equation}}
\newcommand{\ee}{\end{equation}}
\newcommand{\ba}{\begin{eqnarray}}
\newcommand{\ea}{\end{eqnarray}}
\newcommand{\ban}{\begin{eqnarray*}}
\newcommand{\ean}{\end{eqnarray*}}

\begin{document}
                                                                               
\title{Instabilities driven equilibration \\
of the quark-gluon plasma\footnote{Extended version of the review
presented at 18-th International Conference on Nucleus-Nucleus
Collisions {\it Quark Matter 2005}, Budapest, Hungary, August 4-9, 2005.}}
\author{Stanis\l aw Mr\' owczy\' nski\footnote{Electronic address:
{\tt mrow@fuw.edu.pl}}}

\affiliation{ Institute of Physics, \'Swi\c etokrzyska Academy, \\
ul. \'Swi\c etokrzyska 15, PL - 25-406 Kielce, Poland \\
and So\l tan Institute for Nuclear Studies, \\
ul. Ho\.za 69, PL - 00-681 Warsaw, Poland}

\date{November 28, 2005}

\begin{abstract}

Due to anisotropic momentum distributions the parton system
produced at the early stage of relativistic heavy-ion collisions
is unstable with respect to the magnetic plasma modes. The 
instabilities isotropize the system and thus speed up the 
process of its equilibration. The whole scenario of the 
instabilities driven isotropization is reviewed. 

\end{abstract}

\pacs{12.38.Mh, 25.75.-q}


\maketitle

                                                                                
\section{Introduction}
                                                                                

The matter created in relativistic heavy-ion collisions manifests
a strongly collective hydrodynamic behaviour \cite{Heinz:2005ja}
which is particularly evident in studies of the so-called elliptic 
flow \cite{Retiere:2004wa}. A hydrodynamic description requires, strictly 
speaking, a local thermal equilibrium and experimental data on the 
particle spectra and elliptic flow suggest, when analysed within the
hydrodynamic model, that an equilibration time of the parton\footnote{The 
term `parton' is used to denote a quasiparticle fermionic (quark) or 
bosonic (gluon) excitation of the quark-gluon plasma.} system produced 
at the collision early stage is as short as 0.6 ${\rm fm}/c$ 
\cite{Heinz:2004pj}. Such a fast equilibration can be explained 
assuming that the quark-gluon plasma is strongly coupled 
\cite{Shuryak:2004kh}. However, it is not excluded that due to 
the high-energy density at the early stage of the collision, 
when the elliptic flow is generated \cite{Sorge:1998mk}, the plasma 
is weakly coupled because of asymptotic freedom. Thus, the question 
arises whether the weakly interacting plasma can be 
equilibrated within 1 ${\rm fm}/c$.

Models that assume that parton-parton collisions are responsible 
for the thermalization of weakly coupled plasma provide a 
significantly longer equilibration time. The calculations 
performed within the `bottom-up' thermalization scenario 
\cite{Baier:2000sb}, where the binary and $2 \leftrightarrow 3$
processes are taken into account, give an equilibration time of 
at least 2.6 ${\rm fm}/c$ \cite{Baier:2002bt}. To thermalize the 
system one needs either a few hard collisions of momentum transfer 
of order of the characteristic parton momentum\footnote{Although 
anisotropic systems are considered, the characteristic momentum in 
all directions is assumed to be of the same order.}, which is denoted 
here as $T$ (as the temperature of equilibrium system), or many 
collisions of smaller transfer. As discussed in {\it e.g.} 
\cite{Arnold:1998cy}, the inverse equilibration time is of 
order $g^4 {\rm ln}(1/g)\,T$ (with $g$ being the QCD coupling 
constant) when the binary collisions are responsible for the system's
thermalization. However, the equilibration is speeded up by 
instabilities generated in an anisotropic quark-gluon plasma 
\cite{Mrowczynski:xv,Arnold:2004ti}, as growth of the unstable 
modes is associated with the system's isotropization. The 
characteristic inverse time of instability development is roughly of 
order $gT$ for a sufficiently anisotropic momentum distribution
\cite{Mrowczynski:xv,Arnold:2004ti,Randrup:2003cw,Romatschke:2003ms,Arnold:2003rq,Rebhan:2004ur}.
Thus, the instabilities are much `faster' than the collisions
in the weak coupling regime. Recent numerical simulation 
\cite{Dumitru:2005gp} shows that the instabilities driven 
isotropization is indeed very efficient.
                                                                                
The isotropization should be clearly distinguished from the 
equilibration. The instabilities driven isotropization is 
a mean-field reversible phenomenon which is {\em not} accompanied 
with entropy production \cite{Mrowczynski:xv,Dumitru:2005gp}. 
Therefore, the collisions, which are responsible for the dissipation, 
are needed to reach the equilibrium state of maximal entropy. The 
instabilities contribute to the equilibration indirectly, shaping 
the parton momenta distribution. And recently it has been argued 
\cite{Arnold:2004ti} that the hydrodynamic collective behaviour 
does not actually require local thermodynamic equilibrium but 
a merely isotropic momentum distribution of liquid components. 
Thus, the above mentioned estimate of 0.6 ${\rm fm}/c$
\cite{Heinz:2004pj} rather applies to the isotropization than 
to the equilibration. 

My aim here is to review the whole scenario of instabilities 
driven isotropization and the article is organized as follows. 
I start with a brief presentation of numerous efforts to 
understand the equilibration process of the quark-gluon plasma 
which have been undertaken over last two decades. 
In Sec.~\ref{sec-rele} various plasma instabilities are considered 
and the magnetic Weibel modes are argued to be relevant for the 
quark-gluon plasma produced in relativistic heavy-ion collisions. 
In Sec.~\ref{sec-seeds} I discuss how the unstable modes are 
initiated while in Sec.~\ref{sec-mechanism} the mechanism of 
unstable mode growth is explained in terms of elementary 
physics. Sec.~\ref{sec-dispersion} is devoted to solutions of 
the dispersion equation which provide dispersion relations 
of the unstable modes. In Sec.~\ref{sec-iso-abel} it is explained 
why the instabilities isotropize the system. A phenomenon 
of spontaneous abelianization of the system's configuration 
is considered in the same section. The two next sections contain
more formal material. The Hard Loop effective action 
of anisotropic plasma is presented in Sec.~\ref{sec-eff-action}
while Sec.~\ref{sec-eq-motion} deals with the equations of motion 
which are used to study temporal evolution of anisotropic plasma.
Results of recent numerical simulations of the plasma evolution 
are presented in Sec.~\ref{sec-simulate}. The review is closed 
with a brief discussion on possible signals of the instabilities 
and on desired improvements of theoretical approaches to the 
unstable quark-gluon plasma.

Throughout the article there are used the natural units with 
$\hbar = c = k_{\rm B} = 1$; the metric convention 
is $(1,-1,-1,-1)$; the coupling constant $\alpha_s \equiv g^2/4\pi$ 
is assumed to be small; quarks and gluons are massless.

                                                                                
\section{Equilibration of the Quark-Gluon Plasma}
\label{sec-equi}
                                                                                

To present the scenario of instabilities driven isotropization
in a broader context, I start with a brief review of numerous
attempts to understand the equilibration processes of the
quark-gluon. The problem was posed over twenty years ago when 
the real prospects to create the quark-gluon plasma in terrestrial 
experiments appeared. Already in the early papers published in 
the eighties 
\cite{Baym:1984np,Chakraborty:1985ha,Kajantie:1985jh,Hwa:1985tv,Boal:1986nr,
Eskola:1988hp,Banerjee:1989by}, 
main directions of further studies were drawn. The space-time 
structure of ultrarelativistic heavy-ion collisions was found 
\cite{Hwa:1985tv} to provide an estimate of the system's temperature 
and the lower bound of the thermalization time. The Boltzmann equation 
in the relaxation time approximation \cite{Baym:1984np} and the 
Fokker-Planck equation \cite{Chakraborty:1985ha} were used to 
follow the equilibration process. The Schwinger mechanism of 
particle production was included in kinetic theory treatment of 
the thermalization \cite{Kajantie:1985jh,Banerjee:1989by} and the
pure perturbative mechanism was analysed as well \cite{Eskola:1988hp}. 
The equilibration was also studied within the Monte Carlo parton 
cascade model \cite{Boal:1986nr} which, however, took into account 
only binary parton-parton collisions.  

These lines of research were continued in the next decade. The parton 
cascade approach was greatly improved \cite{Geiger:1991nj} by, in 
particular, including the gluon radiation in the initial and final 
states of parton-parton interactions. The radiation proved to be 
very important for the equilibration process 
\cite{Geiger:1992si,Geiger:1992ac}. These detailed numerical studies 
are summarized in the review \cite{Geiger:1994he}. Another perturbative 
parton cascade approach combined with the string phenomenology for 
non-perturbative interactions is presented in \cite{Wang:1996yf}. 
The analytical studies of the thermalization were continued in 
\cite{Shuryak:1992wc,Biro:1993qt,Alam:1994sc,Heiselberg:1995sh,Heiselberg:1996xg}, 
see also \cite{Wang:1996yf}, where, in particular, the gluons were
convincingly shown to equilibrate much faster than the quarks, 
the free streaming and the role of infrared cut-offs in the 
parton-parton cross sections were elucidated. Much efforts 
were invested in the studies of multi-particle processes 
\cite{Xiong:1992cu,Wong:1996ta,Wong:1996va,Wong:1997dv}
which were already implemented in the parton cascade type
models \cite{Geiger:1994he,Wang:1996yf}. The inelastic process 
$2 \leftrightarrow 3$ attracted a lot of attention. Although it 
is of higher order in $\alpha_s$, it is responsible for the parton
number equilibration and it dominates the entropy production
\cite{Wong:1996ta,Wong:1996va,Wong:1997dv}.

There are two very recent transport theory approaches to the 
equilibration problem based on big numerical codes where the 
role of the multi-particle processes is emphasized 
\cite{Xu:2004mz,Xu:2004gw}. The authors of \cite{Xu:2004mz} 
include particle production and absorption via the process 
$2 \leftrightarrow 3$ while the three-particle collisions
$3 \leftrightarrow 3$ are studied in \cite{Xu:2004gw}. 
Within both approaches the equilibration is claimed to be 
significantly speeded-up when compared to the equilibration 
driven by the binary collisions. However, the interaction rates 
of multi-particle processes are known to suffer from severe
divergences, and thus, the actual role of the multi-particle 
interactions crucially depends on how the rates are defined,
computed and regularized.  

The observation that the multi-particle interaction rates are 
sometimes divergent was actually used to explain the very fast 
equilibration of the quark-gluon system. The so-called collinear 
divergences of the gluon multiplication process $2 \leftrightarrow 3$ 
cancel in the equilibrium. If the cancellation does not occur in 
the non-equilibrium systems, as argued in \cite{Wong:2004ik}, the 
equilibration, which is driven by very large - formally divergent 
- interaction rates, is extremely fast even in the weakly coupled 
plasma \cite{Wong:2004ik}.

The thermalization of the quark-gluon plasma was also discussed
from a very different point of view where the equilibration
is not due to the inter-parton collisions but due to the chaotic 
dynamics of the non-Abelian classical fields (coupled or not 
to the classical coloured particles) 
\cite{Biro:1993qc,Sengupta:1999jy}, see also a very recent paper 
\cite{Bannur:2005wz}. Then, the equilibration time is controlled 
by the maximal Lyapunov exponent. 

At the turn of the millennium, when a large volume of
experimental data from the RHIC started to flow, understanding
of the equilibration process became a burning issue as
the data favoured a very short equilibration mentioned
in the Introduction. Within the concept of strongly coupled
quark-gluon plasma, the problem is trivially solved as 
the strongly interacting system is indeed equilibrated 
very fast. However, it is still an open issue whether the 
plasma at the collision early stage is indeed strongly coupled. 

A novel development concerned a treatment of the initial
state of the parton system which evolves towards equilibrium.
In the papers mentioned above, one usually assumed that the 
initial partons are produced due to the (semi-)hard interactions 
of partons of the incident nuclei. Thus, jets and minijets form 
such an initial state which can be parametrized in several ways 
\cite{Cooper:2002td}. Recent studies of the equilibration 
problem which adopt the minijet initial conditions are presented 
in \cite{ Bhalerao:1999hj,Nayak:2000js,Shin:2003yk}. 

In the already mentioned `bottom-up' thermalization scenario 
\cite{Baier:2000sb}, the initial state was assumed to be 
shaped by the QCD saturation mechanism. Then, the initial 
state is dominated by the small $x$ gluons of transverse
momentum of order $Q_s$ which is the saturation scale.
These gluons are freed from the incoming nuclei after a time 
$~Q_s^{-1}$. Weak coupling techniques are applicable 
as $Q_s$ is expected to be much smaller than $\Lambda_{\rm QCD}$ 
at sufficiently high collision energies. The saturation mechanism
is incorporated in the effective field approach known as the 
Colour Glass Condensate \cite{Iancu:2003xm} where the small $x$ 
partons of large occupation numbers are treated as classical 
Yang-Mills fields. Hard modes of the classical fields play the 
role of particles here. The equilibration processes with the
minijet and saturation initial states were compared to each
other in \cite{Serreau:2001xq}.

The `bottom-up' thermalization scenario \cite{Baier:2000sb},
where not only binary collisions but the processes 
$2 \leftrightarrow 3$ are included, takes into account 
the system's expansion. The equilibration processes splits into
several stages parametrically characterized by $\alpha_s^n Q_s^{-1}$
where $n$ is a fractional power. The thermalization time 
is of order $\alpha_s^{-13/5}Q_s^{-1}$. However, as stressed
in the Introduction, the collisional isotropization is apparently 
too slow to comply with the experimental data. The calculations 
performed within the `bottom-up' scenario \cite{Baier:2000sb} 
were criticized \cite{Arnold:2003rq} for treating the parton 
momentum distribution as isotropic, and thus, ignoring the 
instabilities which actually speed up the equilibration process. 
Recently, an influence of the instabilities on the `bottom-up' 
time scales has been discussed in \cite{Bodeker:2005nv}. It has 
been also argued \cite{Mueller:2005un} that a somewhat modified 
scenario remains valid for a sufficiently late stage of the 
equilibration process when the instabilities are no longer 
operative. 

At the end I mention rather unconventional approaches to
the fast equilibration problem. It was argued in
\cite{Bialas:1999zg,Florkowski:2003mm,Kharzeev:2005iz}
that the momentum distribution of partons is of the equilibrium 
form just after the production process. Thus, the very process 
of particle production leads to the equilibrium state without 
any secondary interactions. The authors of 
\cite{Bialas:1999zg,Florkowski:2003mm} refer to the Schwinger
mechanism of particle's production due to the strong chromoelectric 
field. The transverse momentum but not longitudinal one is 
claimed to be `equilibrated' in this way 
\cite{Bialas:1999zg,Florkowski:2003mm}. The key ingredient 
of the approach \cite{Kharzeev:2005iz}, where the longitudinal
momentum is also thermal, is the Hawking-Unruh 
effect: an observer moving with an acceleration $a$ experiences 
the influence of a thermal bath with an effective temperature 
$a / 2\pi$, similar to the one present in the vicinity of 
a black hole horizon. The idea behind the approaches 
\cite{Bialas:1999zg,Florkowski:2003mm,Kharzeev:2005iz}
is elegant and universal -- it can be applied not only
to nucleus-nucleus but to hadron-hadron or even to $e^+\!-\!e^-$ 
collisions -- but it cannot explain how the equilibrium state 
is maintained when the parton's free streaming drives the 
system out of equilibrium. Secondary interactions are
then certainly needed.

Finally, I note a very interesting `no-go' theorem 
\cite{Kovchegov:2005ss,Kovchegov:2005kn}, which states that 
the perturbative thermalization is impossible, as any Feynman 
diagram of any order leads in the long time limit to the time 
scaling of the energy density corresponding to the free streaming 
not to the Bjorken hydrodynamics. However, it is not quite clear 
whether the theorem applies to the relativistic heavy-ion collisions 
as the equilibrium state of matter produced in the collisions is 
presumably only a transient state which changes into free 
streaming at the late times of the system's evolution. 

                                                                                
\section{Relevant Plasma Instabilities}
\label{sec-rele}
                                                                                

The electron-ion plasma is known to experience a large variety of 
instabilities \cite{Kra73}. Those caused by coordinate space 
inhomogeneities, in particular by the system's boundaries, are 
usually called {\it hydrodynamic} instabilities, while those due 
to non-equilibrium momentum distribution of plasma particles 
are called {\it kinetic} instabilities. Hardly anything is known 
about hydrodynamic instabilities of the quark-gluon plasma, and 
I will not speculate about their possible role in the system's 
dynamics. The kinetic instabilities are initiated either by 
the charge or current fluctuations. In the first case, the electric 
field (${\bf E}$) is longitudinal (${\bf E} \parallel {\bf k}$, where 
${\bf k}$ is the wave vector), while in the second case the field 
is transverse (${\bf E} \perp {\bf k}$). For this reason, the kinetic 
instabilities caused by the charge fluctuations are usually called 
{\it longitudinal} while those caused by the current fluctuations 
are called {\it transverse}. Since the electric field plays a crucial 
role in the longitudinal mode generation, the longitudinal instabilities 
are also called {\it electric} while the transverse ones are called
{\it magnetic}. In the non-relativistic plasma the electric instabilities 
are usually much more important than the magnetic ones, as the magnetic 
effects are suppressed by the factor $v^2/c^2$ where $v$ is the particle's
velocity. In the relativistic plasma both types of instabilities are 
of similar strength. The electric instabilities occur when the momentum 
distribution of plasma particles has more than one maximum, as in the 
two-stream system. A sufficient condition for the magnetic instabilities 
is, as discussed in Sec.~\ref{sec-dispersion}, anisotropy of the momentum 
distribution. 

Soon after the concept of quark-gluon plasma had been established, 
the existence of the colour kinetic instabilities, fully analogous to 
those known in the electrodynamic plasma, was suggested 
\cite{Heinz:1985vf,Pokrovsky:1988bm,Pokrovsky:1990sz,Pokrovsky:1990uh,Mrowczynski:1988dz,Pavlenko:1990as,Pavlenko:1991ih}. 
In these early papers, however, there was considered a two-stream system, 
or more generally, a momentum distribution with more than one maximum. 
While such a distribution is common in the electron-ion plasma, 
it is rather irrelevant for the quark-gluon plasma produced in relativistic
heavy-ion collisions where the global as well as local momentum distribution
is expected to monotonously decrease in every direction from the maximum.
The electric instabilities are absent in such a system but, as demonstrated
in \cite{Mrowczynski:1993qm,Mrowczynski:xv}, a magnetic unstable mode 
known as the filamentation or Weibel instability \cite{Wei59} is possible.
The filamentaion instability was shown \cite{Mrowczynski:1993qm,Mrowczynski:xv}
to be relevant for the quark-gluon plasma produced in relativistic 
heavy-ion collisions as the characteristic time of instability growth 
is shorter or at least comparable to other time scales of the parton 
system evolution. And the instabilities -- usually not one but several
modes are generated -- drive the system towards isotropy, thus speeding 
up its equilibration. In the following sections a whole scenario of
the instabilities driven equilibration is reviewed.

                                                                                
\section{Seeds of filamentation}
\label{sec-seeds}
  

Let me start with a few remarks on degrees of freedom of the
quark-gluon plasma. Various problems will be repeatedly discussed
in terms of {\em classical fields} and {\em particles} which are 
only approximate notions in the quark-gluon plasma being a system 
of relativistic quantum fields. However, collective excitations, 
which are bosonic and highly populated, can be treated as classical 
fields while bosonic or fermionic excitations, with the energy 
determined by the excitation momentum (due to the dispersion relation), 
can be treated as (quasi-)particles. In the weakly coupled quark-gluon 
plasma in equilibrium, an excitation is called {\em hard} when its momentum 
is of order $T$, which is the system's temperature, and it is called 
{\em soft} when its momentum is of order $gT$. Within the Hard Loop 
dynamics, the hard excitations can be treated as particles while the 
gluonic soft excitations as classical fields \cite{Blaizot:2001nr}.
It is expected that a similar treatment is possible in the non-equilibrium
plasma as well. Thus, the terms partons, quarks, gluons, particles will 
be used to denote quasiparticle hard excitations. The classical 
chromodynamic field will represent gluonic soft collective excitations.

After the introductory remarks, let me discuss how the unstable 
transverse modes are initiated. For this purpose I consider a parton 
system which is homogeneous but the parton momentum distribution is 
not of the equilibrium form, it is {\em not} isotropic. The system 
is on average locally colourless but colour fluctuations are possible. 
Therefore, 
$\langle j^{\mu}_a (x)\rangle = 0$ where $j^{\mu}_a (x)$ is a local
colour four-current in the adjoint representation of ${\rm SU}(N_c)$ 
gauge group with $\mu=0,1,2,3$ and $a =1,2, \dots , (N_c^2 -1)$ being 
the Lorentz and colour index, respectively; $x=(t,{\bf x})$ denotes 
a four-position in coordinate space.  
 
Since I assume that the quark-gluon plasma is weakly coupled, the
non-interacting gas of quarks, antiquarks and gluons can be treated as
a first approximation. As discussed in detail in \cite{Mrowczynski:1996vh}, 
the current correlator for a classical system of non-interacting 
massless partons is 
\ba
\label{cur-cor-x}
M^{\mu \nu}_{ab} (t,{\bf x}) &\buildrel \rm def \over =& 
\langle j^{\mu}_a (t_1,{\bf x}_1) j^{\nu}_b (t_2,{\bf x}_2) \rangle 
= {1 \over 8} \,g^2\; \delta^{ab} 
\int {d^3p \over (2\pi )^3} \; {p^{\mu} p^{\nu} \over {\bf p}^2} \;
f({\bf p}) \; \delta^{(3)} ({\bf x} -{\bf v} t)  \;,
\ea
where ${\bf v} \equiv {\bf p}/ |{\bf p}|$, 
$(t,{\bf x}) \equiv (t_2-t_1,{\bf x}_2-{\bf x}_1)$ and the 
effective parton distribution function $f({\bf p})$ equals 
$n({\bf p}) + \bar n({\bf p}) + 2N_c n_g({\bf p})$; $n({\bf p})$, 
$\bar n({\bf p})$ and $n_g({\bf p})$ give the average colourless
distribution function of quarks $Q^{ij}({\bf p},x)= \delta^{ij} n({\bf p})$, 
antiquarks $\bar Q^{ij}({\bf p},x)= \delta^{ij} \bar n({\bf p})$, and 
gluons $G^{ab}({\bf p},x)= \delta^{ab}n_g({\bf p})$. The distribution 
function of (anti-)quarks and gluons are matrices belonging to the 
fundamental and ajoint representation, respectively, of the 
${\rm SU}(N_c)$ gauge group. Therefore, $i,j=1,2,\dots , N_c$ and 
$a,b =1,2,..., (N_c^2 -1)$.

Due to the average space-time homogeneity, the correlation tensor 
(\ref{cur-cor-x}) depends only on the difference 
$(t_2-t_1,{\bf x}_2-{\bf x}_1)$. The space-time points $(t_1,{\bf x}_1)$
and $(t_2,{\bf x}_2)$ are correlated in the system of non-interacting 
particles if a particle travels from $(t_1,{\bf x}_1)$ to $(t_2,{\bf x}_2)$. 
For this reason the delta  $\delta^{(3)} ({\bf x} - {\bf v} t)$ is 
present in the formula (\ref{cur-cor-x}). The momentum integral of the
distribution function simply represents the summation over particles.
The fluctuation spectrum is found as a Fourier transform of the tensor 
(\ref{cur-cor-x}) {\it i.e.} 
\be
\label{cur-cor-k}
M^{\mu \nu}_{ab} (\omega ,{\bf k}) = {1 \over 8} \,g^2\; \delta^{ab} 
\int {d^3p \over (2\pi )^3} \; 
{p^{\mu} p^{\nu} \over {\bf p}^2} \; f({\bf p})  \;
2\pi \delta (\omega -{\bf kv}) \;.
\ee

To compute the fluctuation spectrum, the parton momentum distribution
has to be specified. Such calculations with two forms of the  
momentum distribution are presented in \cite{Mrowczynski:1996vh}. Here 
I only qualitatively discuss Eqs.~(\ref{cur-cor-x},~\ref{cur-cor-k}),
assuming that the parton momentum distribution is anisotropic.

In heavy-ion collisions, the anisotropy is a generic feature of the 
parton momentum distribution in a local rest frame. After the first 
collisions, when the partons are released from the incoming nucleons, 
the momentum distribution is strongly elongated along the beam - it 
is of the prolate shape with the average transverse momentum being 
much smaller than the average longitudinal one. Due to the free 
streaming, it evolves in the local rest frame to the distribution 
which is squeezed along the beam - it is of the oblate shape with 
the average transverse momentum being much larger than the average 
longitudinal one. In most cases, I assume that the distribution is 
elongated along the $z$ axis but my considerations remain valid for 
the distribution, which is squeezed along the $z$ axis, but the 
axes should be relabeled.

With the momentum distribution elongated in the  $z$ direction, 
Eqs.~(\ref{cur-cor-x},~\ref{cur-cor-k}) clearly show that the 
correlator $M^{zz}$ is larger than  $M^{xx}$ or  $M^{yy}$. It is also 
clear that $M^{zz}$ is the largest when the wave vector ${\bf k}$ is 
along the direction of the momentum deficit. Then, the delta function 
$\delta (\omega -{\bf kv})$ does not much constrain the integral in 
Eq.~(\ref{cur-cor-k}). Since the momentum distribution is elongated 
in the $z$ direction, the current fluctuations are the largest when 
the wave vector ${\bf k}$ is the $x\!-\!y$ plane. Thus, I conclude 
that some fluctuations in the anisotropic system are large, much larger
than in the isotropic one. An anisotropic system has a natural
tendency to split into the current filaments parallel to the direction 
of the momentum surplus. These currents are seeds of the filamentation
instability.


\section{Mechanism of filamentation}
\label{sec-mechanism}


Let me now explain in terms of elementary physics why the fluctuating 
currents, which flow in the direction of the momentum surplus, can grow
in time. To simplify the discussion, which follows \cite{Mrowczynski:1996vh}, 
I consider an electromagnetic anisotropic system. The form of the fluctuating 
current is chosen to be
\be
\label{flu-cur}
{\bf j}(x) = j \: \hat {\bf e}_z \: {\rm cos}(k_x x) \;,
\ee
where $\hat {\bf e}_z$ is the unit vector in the $z$ direction.
As seen in Eq.~(\ref{flu-cur}), there are current filaments of 
the thickness $\pi /\vert k_x\vert$ with the current flowing in the 
opposite directions in the neighbouring filaments. 

\begin{figure}
\begin{minipage}{19pc}
\includegraphics[width=19.5pc]{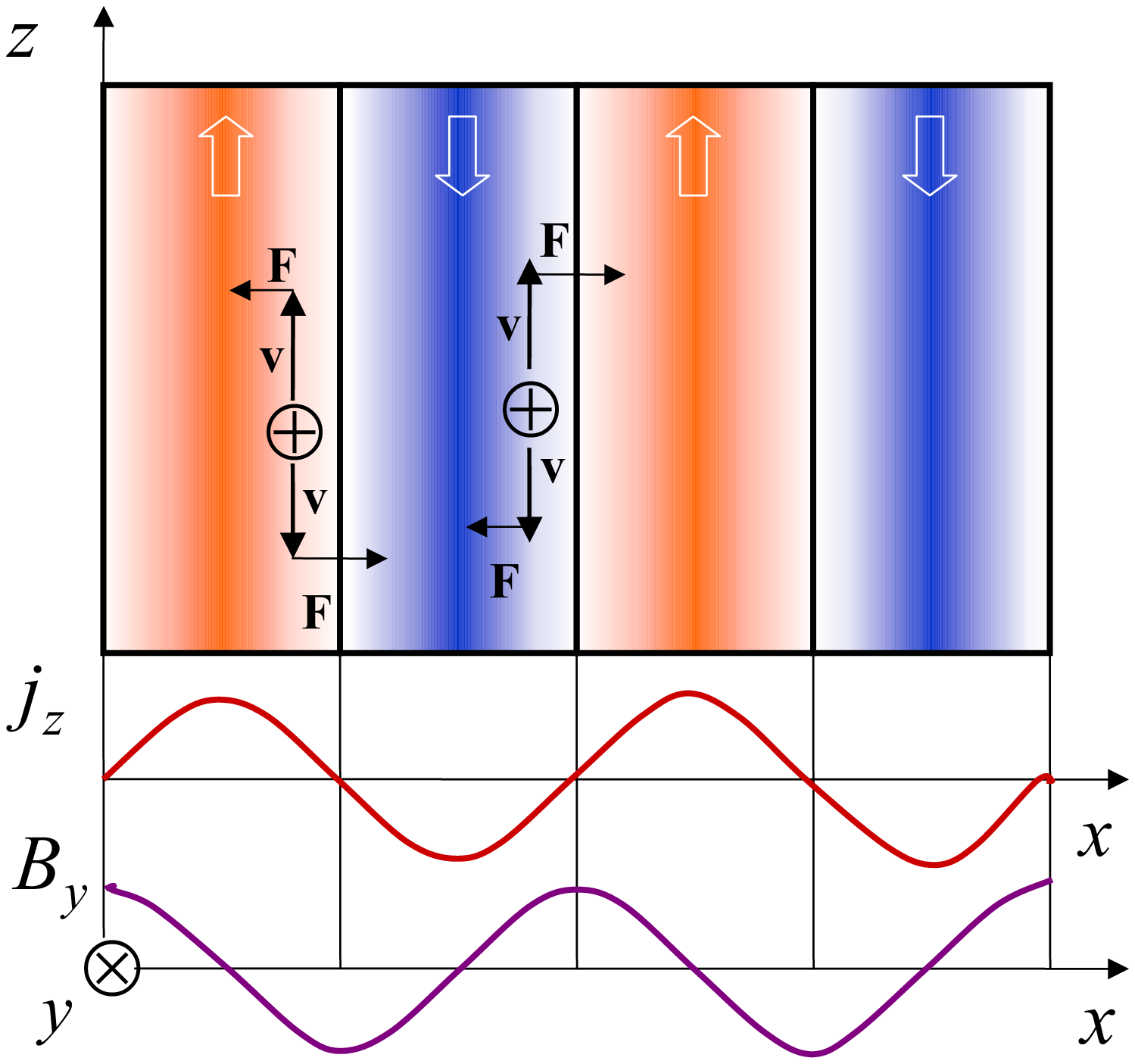}
\vspace{-0.9cm}
\caption{The mechanism of filamentation instability, 
see text for a description.}
\label{fig-mechanism}
\end{minipage}\hspace{2pc}%
\begin{minipage}{16pc}
\vspace{0.7cm}
\includegraphics[width=16pc]{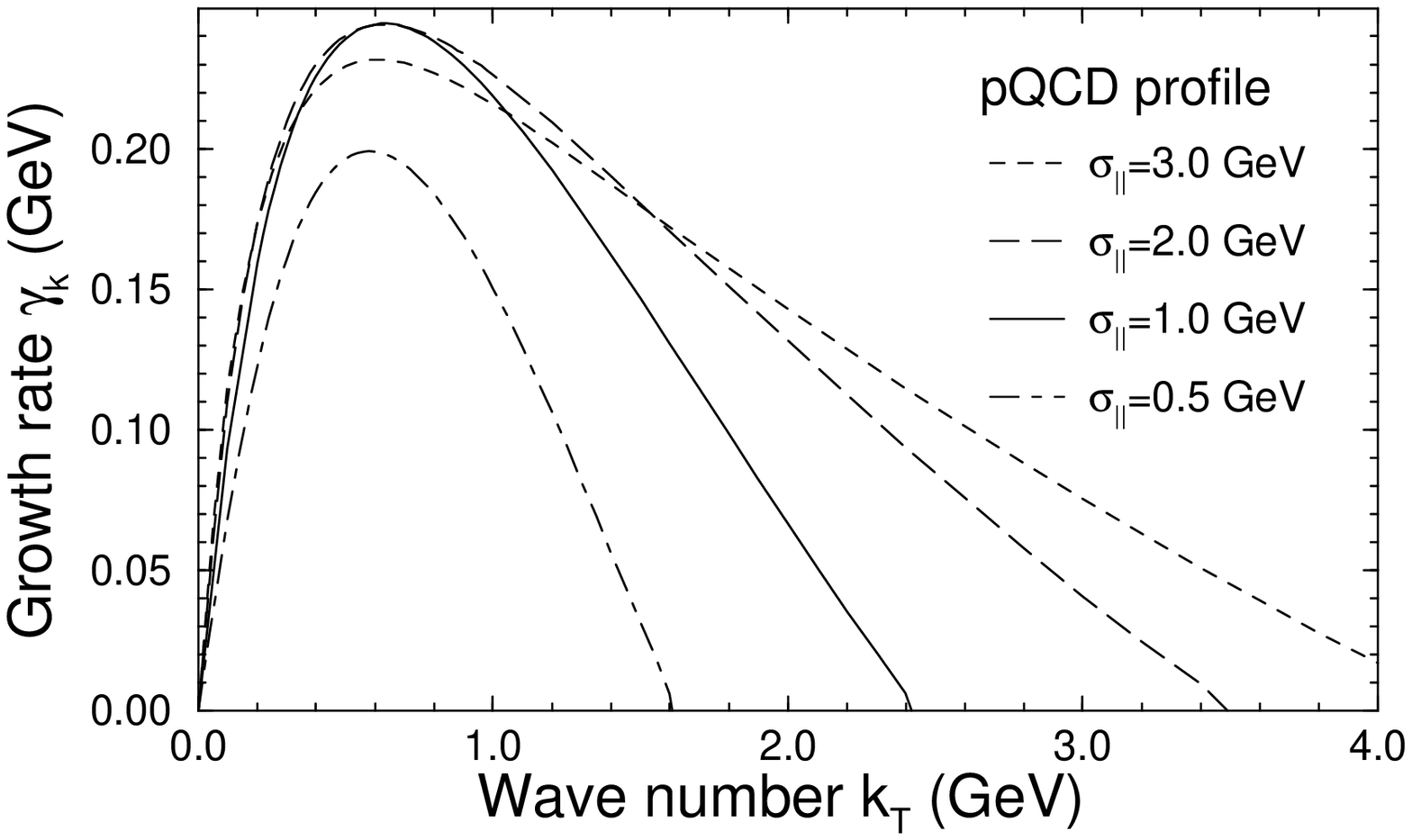}
\vspace{-0.8cm}
\caption{The growth rate of the unstable mode as a function of the
wave vector ${\rm k}=(k_\perp,0,0)$ for $\sigma_\perp=0.3~{\rm GeV}$
and 4 values of the parameter $\sigma_\parallel$ which controls system's 
anisotropy. The figure is taken from \cite{Randrup:2003cw}.}
\label{fig-growth}
\end{minipage}
\end{figure}

The magnetic field generated by the current (\ref{flu-cur}) is given as
\ban
{\bf B}(x) = {j \over k_x} \: \hat {\bf e}_y \: {\rm sin}(k_x x) \;,
\ean
and the Lorentz force acting on the partons, which fly along the $z$
direction, equals
\ban
{\bf F}(x) = q \: {\bf v} \times {\bf B}(x) = 
- q \: v_z \: {j \over k_x} \: \hat {\bf e}_x \: {\rm sin}(k_x x) \;,
\ean
where $q$ is the electric charge. One observes, see Fig.~\ref{fig-mechanism}, 
that the force distributes the partons in such a way that those, which 
positively contribute to the current in a given filament, are focused 
in the filament centre while those, which negatively contribute, are 
moved to the neighbouring one. Thus, the initial current is growing
and the magnetic field generated by this current is growing as well.
The instability is driven by the the energy transferred from the 
particles to fields. More specifically, the kinetic energy related 
to a motion along the direction of the momentum surplus is used to 
generate the magnetic field. The mechanism of Weibel instability is 
explained somewhat differently in \cite{Arnold:2003rq}.


\section{Dispersion equation}
\label{sec-dispersion}


The Fourier transformed chromodynamic field $A^{\mu}(k)$ satisfies the 
equation of motion as
\be
\label{eq-A}
\Big[ k^2 g^{\mu \nu} -k^{\mu} k^{\nu} - \Pi^{\mu \nu}(k) \Big] 
A_{\nu}(k) = 0 \;,
\ee
where $k\equiv (\omega, {\bf k})$ and $\Pi^{\mu \nu}(k)$ is the 
polarization tensor or gluon self-energy which is discussed later 
on. Since the tensor is proportional to a unit matrix in the colour 
space, the colour indices are dropped here. A general plasmon 
dispersion equation is of the form 
\be
\label{dispersion-pi}
{\rm det}\Big[ k^2 g^{\mu \nu} -k^{\mu} k^{\nu} - \Pi^{\mu \nu}(k) \Big] 
 = 0 \;.
\ee
Equivalently, the dispersion relations are given by the positions of 
poles of the effective gluon propagator. Due to the transversality of 
$\Pi^{\mu \nu}(k)$ ($k_\mu \Pi^{\mu \nu}(k) = k_\nu \Pi^{\mu \nu}(k)=0$) 
not all components of $\Pi^{\mu \nu}(k)$ are independent from each other, 
and consequently the dispersion equation (\ref{dispersion-pi}), which 
involves a determinant of a $4\times4$ matrix, can be simplified to the
determinant of a $3\times3$ matrix. For this purpose I introduce the 
colour permittivity tensor $\epsilon^{lm}(k)$ where the indices 
$l,m,n = 1,2,3$ label three-vector and tensor components. Because of 
the relation 
\ban
\epsilon^{lm}(k) E^l(k) E^m(k) = \Pi^{\mu \nu} (k) A_{\mu}(k) A_{\nu}(k)\;,
\ean
where ${\bf E}$ is the chromoelectric vector, the permittivity can be 
expressed through the polarization tensor as
\ban
\epsilon^{lm}(k) = \delta^{lm} + {1 \over \omega^2} \Pi^{lm}(k) \;.
\ean
Then, the dispersion equation gets the form
\be
\label{dispersion-g}
{\rm det}\Big[ {\bf k}^2 \delta^{lm} -k^l  k^m 
- \omega^2 \epsilon^{lm}(k)  \Big]  = 0 \,.
\ee
The relationship between Eq.~(\ref{dispersion-pi}) and 
Eq.~(\ref{dispersion-g}) is most easily seen in the Coulomb 
gauge when $A^0 = 0$ and ${\bf k} \cdot {\bf A}(k)=0$. Then, 
${\bf E} = i\omega {\bf A}$ and Eq.~(\ref{eq-A}) is immediately 
transformed into an equation of motion of ${\bf E}(k)$ which
further provides the dispersion equation (\ref{dispersion-g}).

The dynamical information is contained in the polarization tensor 
$\Pi^{\mu \nu}(k)$ given by Eq.~(\ref{g-self}) or, equivalently, 
in the permittivity tensor $\epsilon^{lm}(k)$ which can be derived 
either within the transport theory or diagrammatically 
\cite{Mrowczynski:2000ed}. The result is
\be
\label{epsilon}
\epsilon^{nm} (\omega, {\bf k}) = \delta^{nm} + 
{g^2 \over 2\omega} \int {d^3 p \over (2\pi )^3}
{ v^n \over \omega - {\bf k v} + i0^+} 
{\partial f({\bf p}) \over \partial p^l} 
\Bigg[ \Big( 1 - {{\bf k v} \over \omega} \Big) \delta^{lm}
+ {k^l v^m \over \omega} \Bigg] \,.
\ee
As already mentioned, the colour indices are suppressed here.

Substituting the permittivity (\ref{epsilon}) into 
Eq.~(\ref{dispersion-g}), one fully specifies the dispersion
equation (\ref{dispersion-g}) which provides a spectrum of
quasi-particle bosonic excitations. A solution $\omega ({\bf k})$ 
of Eq.~(\ref{dispersion-g}) is called {\it stable} when 
${\rm Im}\,\omega \le 0$ and {\it unstable} when ${\rm Im}\,\omega > 0$. 
In the first case the amplitude is constant or it exponentially 
decreases in time while in the second one there is an exponential 
growth of the amplitude. In practice, it appears difficult to find 
solutions of Eq.~(\ref{dispersion-g}) because of the rather complicated 
structure of the tensor (\ref{epsilon}). However, the problem 
simplifies as we are interested in specific modes which are expected 
to be unstable. Namely, we look for solutions corresponding 
to the fluctuating current in the direction of the momentum surplus 
and the wave vector perpendicular to it. 

As previously, the momentum distribution is assumed to be elongated
in the $z$ direction, and consequently the fluctuating current also 
flows in this direction. The magnetic field has a non-vanishing 
component along the $y$ direction and the electric field in the $z$ 
direction. Finally, the wave vector is parallel to the axis $x$, 
see Fig.~\ref{fig-mechanism}. It is also assumed that the momentum 
distribution obeys the mirror symmetry $f(-{\bf p}) = f({\bf p})$, 
and then the permittivity tensor has only non-vanishing diagonal 
components. Taking into account all these conditions, one simplifies 
the dispersion equation (\ref{dispersion-g}) to the form
\be
\label{H-eq}
H(\omega) \equiv k^2_x - \omega^2 \epsilon^{zz}(\omega, k_x) = 0 \;,
\ee
where only one diagonal component of the dielectric tensor enters.

It appears that the existence of unstable solutions of Eq.~(\ref{H-eq}) 
can be proved without solving it. The so-called Penrose criterion 
\cite{Kra73}, which follows from analytic properties of the 
permittivity as a function of $\omega$, states that {\em the 
dispersion equation $H(\omega ) = 0$ has unstable solutions if} 
$H(\omega = 0) < 0$. The Penrose criterion was applied to the
equation (\ref{H-eq}) in \cite{Mrowczynski:xv} but a much more 
general discussion of the instability condition is presented in 
\cite{Arnold:2003rq}. Not entering into details, there exist
unstable modes if the momentum distribution averaged (with
a proper weight) over momentum length is anisotropic.

To solve the dispersion equation (\ref{H-eq}), the parton 
momentum distribution has to be specified. Several analytic
(usually approximate) solution of the dispersion equation with 
various momentum distributions can be found in 
\cite{Mrowczynski:xv,Romatschke:2003ms,Romatschke:2004jh,Arnold:2003rq}.
A typical example of the numerical solution, which gives the 
unstable mode frequency in the full range of wave vectors is 
shown in Fig.~\ref{fig-growth} taken from \cite{Randrup:2003cw}. 
The momentum distribution is of the form
\ban
f({\bf p}) \sim 
\frac{1}{(p_T^2 + \sigma_\perp^2)^3}\, 
{\rm e}^{-{p_z^2 \over 2\sigma_\parallel^2}}\;,
\ean
where $p_\perp \equiv \sqrt{p_x^2 + p_y^2}$. The mode is pure 
imaginary and $\gamma_k \equiv {\rm Im}\,\omega (k_\perp)$. The 
value of the coupling is $\alpha_s \equiv g^2/4\pi =0.3$, 
$\sigma_\perp = 0.3\; {\rm GeV}$ and the effective parton 
density is chosen to be $6 \; {\rm fm}^{-3}$. As seen, there is 
a finite interval of wave vectors for which the unstable modes exist.

The dispersion equation (\ref{H-eq}) corresponds to a simple
configuration where the wave vector is parallel to the axis $x$ 
and it points to the direction of the momentum deficit while 
the chromoelectric field is parallel to the axis $z$ and it 
points to the momentum surplus. However, there are more general 
unstable modes which are not aligned along the symmetry axes 
of the momentum distribution of particles. The wave vectors 
${\bf k}$ and chromoelectric fields ${\bf E}$ of these modes 
have non-vanishing components in the directions of the momentum 
deficit and momentum surplus, respectively, and ${\bf E}$ is no 
longer perpendicular to ${\bf k}$. Such unstable modes are 
discussed in \cite{Randrup:2003cw}. A quite general analysis 
of the dispersion equation of anisotropic systems is given in 
\cite{Romatschke:2003ms,Romatschke:2004jh}. There is 
considered a class of momentum distributions which can be
expressed as 
\be
\label{Mike-ansatz}
f({\bf p}) = 
f_{\rm iso}(\sqrt{{\bf p}^2 + \xi ({\bf n}{\bf p})^2} ) \;,
\ee
where $f_{\rm iso}(| {\bf p}| )$ is an arbitrary (isotropic) 
distribution, the unit vector ${\bf n}$ defines a preferred 
direction and the parameter $\xi \in (-1,\infty)$ controls 
the magnitude of anisotropy. 

As explained above, the existence of the unstable gluonic modes 
is a generic feature of the anisotropic plasma - even a weak
anisotropy generates the instability. In contrary, the quark 
modes seem to be always stable. Although, a general
proof of the quark mode stability is lacking, the modes appear
to be stable even in the case of extremely anisotropic parton
momentum distribution as in the two-stream system
\cite{Mrowczynski:2001az}. Presumably, the quark modes are 
always stable because their population is constrained by 
Pauli blocking \cite{Mike04}.
 
                                                                                
\section{Isotropization and Abelianization}
\label{sec-iso-abel}
                                                                                

When the instabilities grow the system becomes more isotropic 
because the Lorentz force changes the particle's momenta and 
the growing fields carry an extra momentum. To explain 
the mechanism I assume, as previously, that initially 
there is a momentum surplus in the $z$ direction. The fluctuating 
current flows in the $z$ direction with the wave vector 
pointing in the $x$ direction. Since the magnetic field has a $y$ 
component, the Lorentz force, which acts on partons flying along 
the $z$ axis, pushes the partons in the $x$ direction where there 
is a momentum deficit. Numerical simulations discussed in 
Sec.~\ref{sec-simulate} show that growth of the instabilities is 
indeed accompanied with the system's fast isotropization. 

\begin{figure}
\begin{center}
\includegraphics[width=13cm]{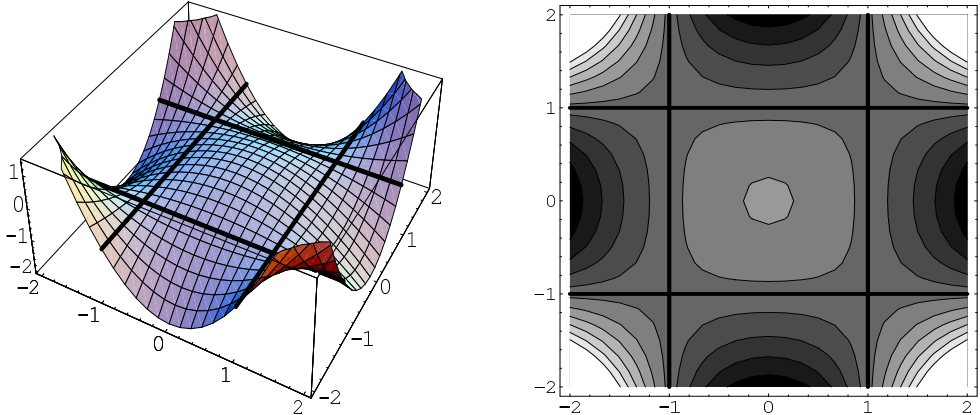}
\end{center}
\vspace{-0.5cm}
\caption{The effective potential of the unstable magnetic mode as
a function of magnitude of two colour components of ${\bf A}^a$ 
belonging to the SU(2) gauge group. The figure is taken from 
\cite{Arnold:2004ih}.}
\label{fig-eff-Abel}
\end{figure}

The system isotropizes not only due to the effect of the Lorentz
force but also due to the momentum carried by the growing field. 
When the magnetic and electric fields are oriented along the
$y$ and $z$ axes, respectively, the Poynting vector points in
the direction $x$ that is along the wave vector. Thus, the momentum 
carried by the fields is oriented in the direction of the momentum 
deficit of particles.

Unstable modes cannot grow to infinity and even in the electron-ion 
plasma there are several possible mechanisms which stop the instability 
growth \cite{Kato:2005wv}. The actual mechanism depends on the 
plasma state as well as on the external conditions. In the case 
of the quark-gluon plasma one suspects that non-Abelian non-linearities 
can play an important role here. An elegant argument \cite{Arnold:2004ih} 
suggests that the non-linearities do not stabilize the unstable modes 
because the system spontaneously chooses an Abelian configuration in 
the course of instability development. Let me explain the idea.

In the Coulomb gauge the effective potential of the unstable 
configuration has the form
\ban
V_{\rm eff}[{\bf A}^a] = - \mu^2 {\bf A}^a \cdot {\bf A}^a
+  \frac{1}{4} g^2 f^{abc} f^{ade}
({\bf A}^b {\bf A}^d) ({\bf A}^c {\bf A}^e) \;,
\ean
which is shown in Fig.~\ref{fig-eff-Abel} taken from \cite{Arnold:2004ih}. 
The first term (with $ \mu^2 >0$) is responsible for the very existence 
of the instability. The second term, which comes from the Yang-Mills
lagrangian, is of pure non-Abelian nature. The term appears to be positive
and thus it counteracts the instability growth. However, the non-Abelian
term vanishes when the potential ${\bf A}^a$ is effectively Abelian, and
consequently, such a configuration corresponds to the steepest decrease 
of the effective potential. Thus, the system spontaneously abelianizes 
in the course of instability growth. In Sec.~\ref{sec-simulate}, where 
the results of numerical simulations are presented, the abelianization 
is further discussed.

                                                                                
\section{Hard-Loop Effective Action}
\label{sec-eff-action}
                                                                                

Knowledge of the gluon polarization tensor or, equivalently, 
the chromoelectric permittivity tensor is sufficient to discuss 
the system's stability and the dispersion relations of unstable 
modes. For more detailed dynamical studies the effective action 
of anisotropic quark-gluon plasma is needed. Such an action for 
a system, which is on average locally colour neutral, stationary and 
homogeneous, was derived in \cite{Mrowczynski:2004kv}. The starting 
point was the effective action which describes an interaction of 
classical fields with currents induced by these fields in the plasma. 
The lagrangian density is quadratic in the gluon and quark fields and 
it equals
\be 
\label{action-2}
{\cal L}_2(x) =  
-\int d^4 y \bigg( {1\over2} A^a_\mu(x) \Pi^{\mu \nu}_{ab}(x-y) A^b_\nu(y) 
+ \bar{\Psi}(x) \Sigma (x-y) \Psi (y) \bigg) \;;
\ee
the Fourier transformed gluon polarization tensor $\Pi^{\mu \nu}_{ab}(k)$ 
and the quark self-energy $\Sigma (k)$ read
\ba 
\label{g-self}
\Pi^{\mu \nu}_{ab}(k) &=& \delta_{ab} { g^2 \over 2 } 
\int {d^3p \over (2\pi)^3} 
{ f({\bf p}) \over |{\bf p}| } 
{ (p\cdot k)(k^\mu p^\nu + p^\mu k^\nu) - k^2 p^{\mu} p^{\nu}
- (p\cdot k)^2 g^{\mu\nu} \over(p\cdot k)^2} \;,
\\ [2mm] 
\label{q-self}
\Sigma(k) &=& g^2 {N_c^2 -1 \over 8 N_c} 
\int {d^3p \over (2\pi)^3} 
{ \tilde f ({\bf p}) \over |{\bf p}|} 
{p \cdot \gamma \over p\cdot k} \;, 
\ea
where $f({\bf p})$ and $\tilde f ({\bf p})$ are the effective
parton distribution functions defined as 
$f({\bf p}) \equiv n({\bf p})+\bar n ({\bf p}) + 
2N_c n_g({\bf p})$ and $\tilde f ({\bf p}) \equiv 
n({\bf p}) + \bar n ({\bf p}) + 2 n_g({\bf p})$; 
$n({\bf p})$, $\bar n ({\bf p})$ and $n_g({\bf p})$ are,
as already mentioned below Eq.~(\ref{cur-cor-x}),
the distribution functions of quarks, antiquarks and gluons 
of single colour component in a homogeneous and stationary 
plasma which is locally and globally colourless; the spin 
and flavour are treated as parton internal degrees of freedom. 
The quarks and gluons are assumed to be massless. The 
polarization tensor (\ref{g-self}) can be derived within 
the semiclassical transport theory 
\cite{Mrowczynski:2000ed,Romatschke:2003ms} or diagrammatically 
\cite{Mrowczynski:2000ed}, following the formal rules of the 
Hard Thermal Loop approach. The quark self-energy (\ref{q-self}) 
has been derived so far only diagrammatically
\cite{Mrowczynski:2000ed,Arnold:2002zm} but the derivation 
is also possible within the transport theory, as it has been done 
in \cite{Blaizot:1993be} for the case of equilibrium plasma. 
The action (\ref{action-2}) holds under the assumption that
the field amplitude is much smaller than $T/g$ where 
$T$ denotes the characteristic momentum of (hard) partons.

Following Braaten and Pisarski \cite{Braaten:1991gm}, the 
lagrangian (\ref{action-2}) was modified to comply with the 
requirement of gauge invariance. The final result, which 
is non-local but manifestly gauge invariant, is 
\cite{Mrowczynski:2004kv}
\ba
\label{HL-action}
{\cal L}_{\rm HL}(x) =  {g^2\over2} 
\int {d^3p \over (2\pi)^3} 
&\bigg[& f({\bf p)} \;
F_{\mu \nu}^a (x)
\bigg({p^\nu p^\rho \over (p \cdot D)^2} \bigg)_{ab} \;
F_\rho^{\;\;b \,\mu} (x)
\\ [2mm] \nonumber 
&+& i {N_c^2 -1 \over 4N_c}
\tilde f ({\bf p}) \;
 \bar{\Psi}(x) {p \cdot \gamma \over p\cdot D}
\Psi (x) \bigg] \;,
\ea
where $F^{\mu \nu}_a$ is the strength tensor and $D$ denotes 
the covariant derivative. The effective action (\ref{HL-action}) 
generates $n-$point functions which obey the Ward-Takahashi 
identities. For the equilibrium plasma the action (\ref{HL-action}) 
is equivalent to that one derived in \cite{Taylor:ia} and in the 
explicitly gauge invariant form in \cite{Braaten:1991gm}. The 
equilibrium Hard Loop action was also found within the semiclassical 
kinetic theory \cite{Blaizot:1993be,Kelly:1994dh}.


\section{Equations of motion}
\label{sec-eq-motion}
                                                                                

Transport theory provides a natural framework to study temporal 
evolution of non-equilibrium systems and it has been applied to 
the quark-gluon plasma for a long time. The distribution functions 
of quarks $(Q)$, antiquarks $(\bar Q)$, and gluons $(G)$, which are
the $N_c \times N_c$ and $(N_c^2-1) \times (N_c^2-1)$ matrices, 
respectively, satisfy the transport equations of the form 
\cite{Elze:1989un,Mrowczynski:1989np}:
\ba
\label{transport-eq}
p^{\mu} D_{\mu}Q({\bf p},x) + {g \over 2}\: p^{\mu}
\bigg\{ F_{\mu \nu}(x), 
{\partial Q({\bf p},x) \over \partial p_{\nu}} \bigg\} 
&=&  0 \;,
\\ \nonumber
p^{\mu} D_{\mu}\bar Q({\bf p},x) - {g \over 2} \: p^{\mu}
\bigg\{ F_{\mu \nu}(x),
{\partial \bar Q({\bf p},x) \over \partial p_{\nu}} \bigg\}
&=& 0 \;,
\\ \nonumber 
p^{\mu} {\cal D}_{\mu}G({\bf p},x) + {g \over 2} \: p^{\mu}
\bigg\{ {\cal F}_{\mu \nu}(x),
{\partial G({\bf p},x) \over \partial p_{\nu}} \bigg\} 
&=& 0 \;,
\ea
where $\{...,...\}$ denotes the anticommutator; the transport 
equation of (anti-)quarks is written down in the fundamental 
representation while that of gluons in the adjoint one. Since 
the instabilities of interest are very fast, much faster than 
the inter-parton collisions, the collision terms are neglected 
in Eqs.~(\ref{transport-eq}). The gauge field, which enters 
the transport equations (\ref{transport-eq}), is generated 
self-consistently by the quarks and gluons. Thus, the transport
equations (\ref{transport-eq}) should be supplemented by
the Yang-Mills equation
\be
\label{yang-mills}
D_{\mu} F^{\mu \nu}(x) = j^{\nu}(x)\; ,
\ee
where the colour current is given as
\ba
\label{current}
j^{\mu }(x) = -g \int \frac{d^3p}{(2 \pi)^3} \: 
\frac{p^{\mu}}{|{\bf p}|} \: \tau_a 
\Big[ {\rm Tr}\big[\tau_a \big( Q({\bf p},x) 
- \bar Q ({\bf p},x)\big)\big]
+{\rm Tr}\big[T_a G({\bf p},x)\big]\Big] \;,
\ea
with $\tau_a$ and $T_a$ being the SU($N_c$) group generators in 
the fundamental and adjoint representation, respectively. There 
is a version of the equations (\ref{transport-eq},~\ref{yang-mills}) 
where colour charges of partons are treated as a classical variable 
\cite{Heinz:1984yq}. Then, the distribution functions depend 
not only on $x$ and ${\bf p}$ but on the colour variable as well.

When the equations (\ref{transport-eq},~\ref{yang-mills}) are
linearized around the state, which is stationary, homogeneous
and locally colourless, the equations provide the Hard Loop
dynamics encoded in the effective action (\ref{HL-action}).
The equations are of particularly simple and elegant form
when the quark $\delta Q({\bf p},x)$, antiquark 
$\delta \bar Q({\bf p},x)$ and gluon $\delta G({\bf p},x)$ 
deviations from the stationary state described by
$Q^{ij}_0({\bf p})= \delta^{ij} n({\bf p})$, 
$\bar Q^{ij}_0({\bf p})= \delta^{ij} \bar n({\bf p})$, and 
$G^{ab}_0({\bf p})= \delta^{ab}n_g({\bf p})$
are parameterised by the field $W^\mu({\bf v},x)$ through
the relations
$$
\delta Q({\bf p},x) = 
g {\partial n({\bf p}) \over \partial p^{\mu}} 
W^\mu({\bf v},x) \;,\;\;\;\;\;\;\;
\delta \bar Q({\bf p},x) =
-g {\partial \bar n({\bf p}) \over \partial p^{\mu}} 
W^\mu({\bf v},x) \;,
$$
$$
\delta G({\bf p},x) =
g {\partial n_g({\bf p}) \over \partial p^{\mu}} 
T_a {\rm Tr}\big[ \tau_a W^\mu({\bf v},x) \big] \;,
$$
where ${\bf v} \equiv {\bf p}/|{\bf p}|$. Then, instead
of the three transport equations (\ref{transport-eq}) one has
one equation 
\be
\label{W-eq}
v_\mu D^\mu W^\nu({\bf v},x) = - v_\rho F^{\rho \nu}(x)
\ee
while the Yang-Mills equation (\ref{yang-mills}) reads
\be
\label{Y-M-W}
D_{\mu} F^{\mu \nu}(x) = j^{\nu }(x) = -g^2 \int \frac{d^3p}{(2 \pi)^3} \: 
\frac{p^{\mu}}{|{\bf p}|} 
{\partial f ({\bf p}) \over \partial p^{\rho}}
W^\rho({\bf v},x) \;,
\ee
where $ v^\mu = (1,{\bf v})$ and, as previously, 
$f({\bf p}) \equiv n({\bf p})+\bar n ({\bf p}) + 2N_c n_g({\bf p})$. 
In contrast to the effective action (\ref{HL-action}), the 
equations (\ref{W-eq},~\ref{Y-M-W}) are local in coordinate space. 
Therefore, the transport equation (\ref{W-eq}) combined with 
Eq.~(\ref{Y-M-W}) is often called local representation of the 
Hard Loop dynamics. The equations (\ref{W-eq},~\ref{Y-M-W}),
which for the isotropic equilibrium plasma were first 
given in \cite{Blaizot:2001nr}, are used in the numerical 
simulations \cite{Rebhan:2004ur,Arnold:2005vb,Rebhan:2005re}
discussed in the next section.

                                                                                
\section{Numerical simulations}
\label{sec-simulate}
                                                                                

Temporal evolution of the anisotropic quark-gluon plasma has 
been recently studied by means of numerical simulations 
\cite{Rebhan:2004ur,Dumitru:2005gp,Arnold:2005vb,Rebhan:2005re}.
The simulations, which have been performed in two very different 
dynamical schemes by three groups of authors, are of crucial 
importance as they convincingly demonstrate a key role of the 
instabilities in the evolution of anisotropic quark-gluon 
plasma.

\begin{figure}
\begin{minipage}{20pc}
\includegraphics[width=20pc]{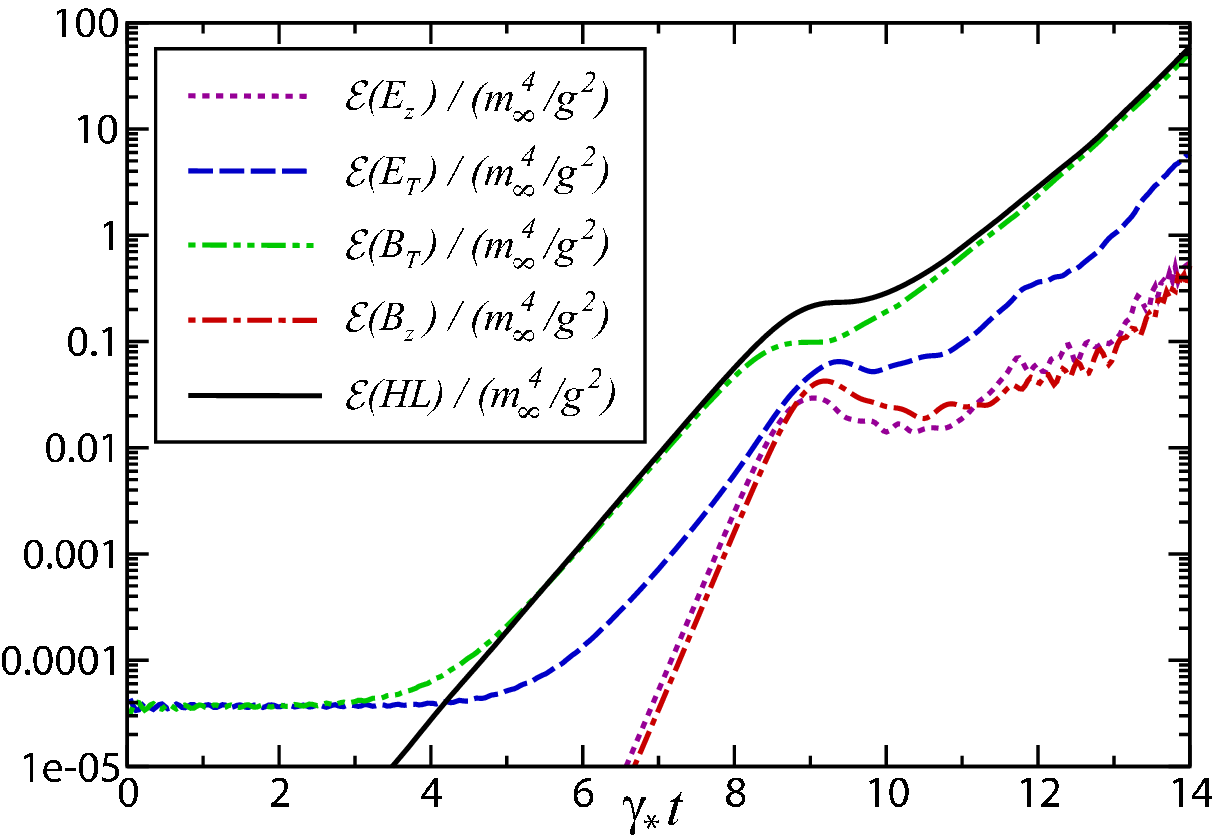}
\vspace{-0.2cm}
\caption{Time evolution of the (scaled) energy density (split into
various electric and magnetic components) which is carried by the 
chromodynamic field. The simulation is 1+1 dimensional and the
gauge group is SU(2). The parton momentum distribution is squeezed 
along the $z$ axis. The solid line corresponds to the total energy
transferred from the particles. The figure is taken from 
\cite{Rebhan:2004ur}.}
\label{fig-energy-rebhan11}
\end{minipage}\hspace{2pc}%
\begin{minipage}{16pc}
\includegraphics[width=16pc]{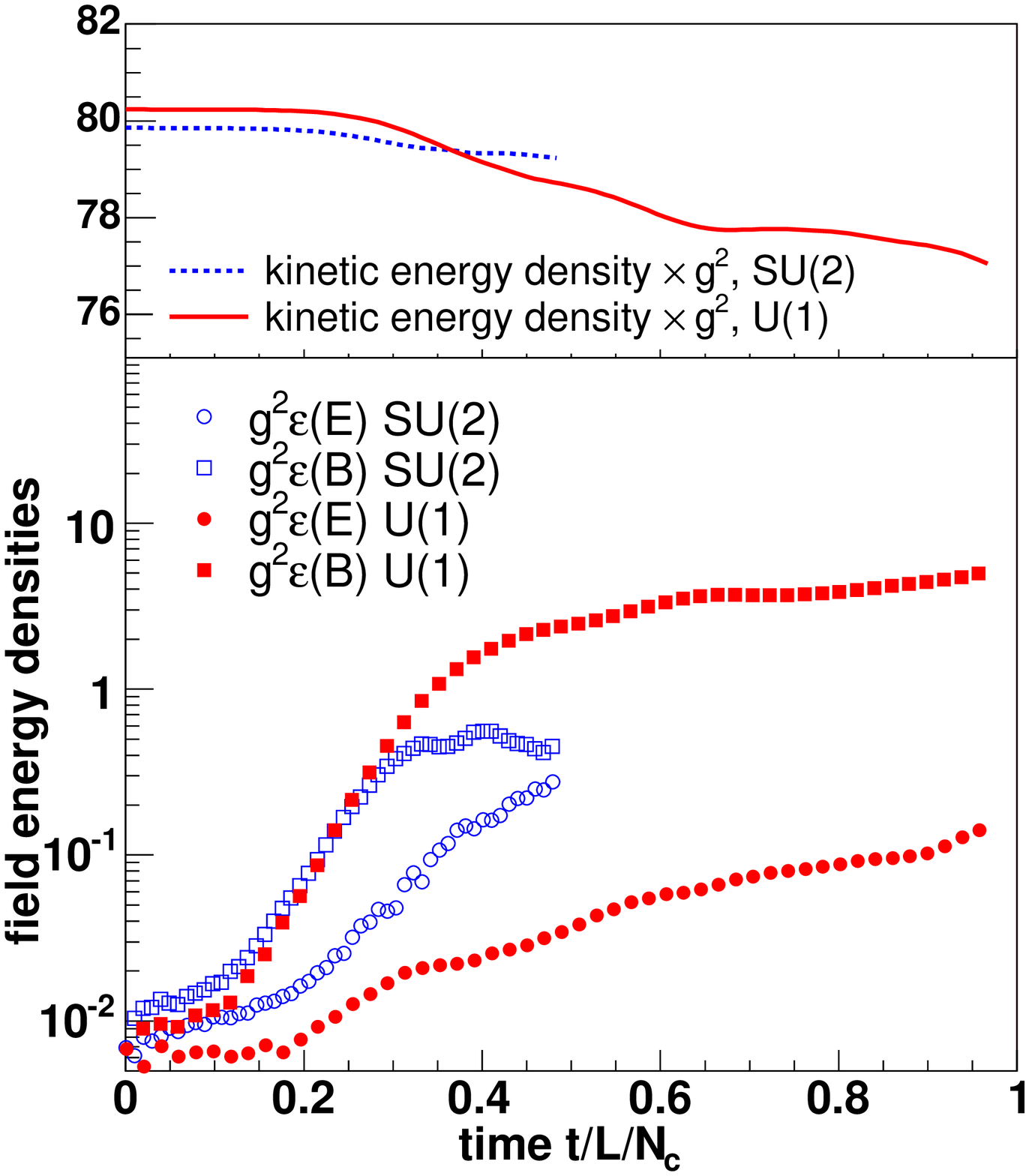}
\vspace{-0.8cm}
\caption{Time evolution of the kinetic energy of particles (upper panel)
and of the energy of electric and magnetic fields (lower panel)
in ${\rm GeV}/{\rm fm}^3$ for the U(1) and SU(2) gauge groups. The 
figure is taken from \cite{Dumitru:2005gp}.}
\label{fig-energy-dumitru}
\end{minipage}
\end{figure}

The dynamics governed by the Hard Loop action (\ref{HL-action}) 
and described by the equations (\ref{W-eq},~\ref{Y-M-W}) 
has been simulated in \cite{Rebhan:2004ur,Arnold:2005vb,Rebhan:2005re}.
These simulations provide fully a reliable information on the field
dynamics provided the potential's amplitude is not too large: 
$A^\mu_a \ll T/g$ where $T$ is the characteristic momentum of 
(hard) partons. Since the equations (\ref{W-eq},~\ref{Y-M-W}) 
describe small deviations from the stationary homogeneous state, 
only a small fraction of the particles is influenced by the 
growing chromodynamic fields. Therefore, the (hard) particles 
effectively play a role of the stationary (anisotropic) background. 
In the simulation \cite{Dumitru:2005gp} the classical version of 
the equations (\ref{transport-eq},~\ref{yang-mills}) is used. The 
quark-gluon plasma is treated as a completely classical system: 
partons, which carry classical colour charges, interact with the 
self-consistently generated classical chromodynamic field.

The simulations \cite{Rebhan:2004ur,Dumitru:2005gp} have been 
effectively performed in 1+1 dimensions as the chromodynamic 
potentials depend on time and one space variable. The calculations 
\cite{Arnold:2005vb,Rebhan:2005re} represent full 1+3 
dimensional dynamics. In most cases the SU(2) gauge group 
was studied but some SU(3) results, which are qualitatively
very similar to SU(2) ones, are given in \cite{Rebhan:2005re}.

The techniques of discretization used in the simulations 
\cite{Rebhan:2004ur,Dumitru:2005gp,Arnold:2005vb,Rebhan:2005re} 
are rather different while the initial conditions are quite similar. 
The initial field amplitudes are distributed according to the 
Gaussian white noise and the momentum distribution of (hard) 
partons is strongly anisotropic. For example, in the classical 
simulation \cite{Dumitru:2005gp} the initial parton momentum 
distribution is chosen as
\be
\label{initial}
f({\bf p}) \sim
\delta(p_x) \;
{\rm e}^{-\frac{\sqrt{p_y^2 + p_z^2}}{p_{\rm hard}}}\;,
\ee
with $p_{\rm hard}= 10 \; {\rm GeV}$.
The results are actually insensitive to the specific form of 
the momentum distribution. If the parton distribution function 
is written in the form (\ref{Mike-ansatz}), the results are shown 
\cite{Rebhan:2004ur,Romatschke:2003ms} to depend only on two 
parameters: $\xi$ and the Debye mass $m_{\rm D}$ of the 
corresponding isotropic system {\it i.e.}
\ban
m_{\rm D}^2 = -{g^2 \over 4\pi^2}
\int_0^{\infty}dp \, p^2
{\partial f_{\rm iso}(p) \over \partial p} \,.
\ean

\begin{figure}
\begin{minipage}{18pc}
\includegraphics[width=17pc]{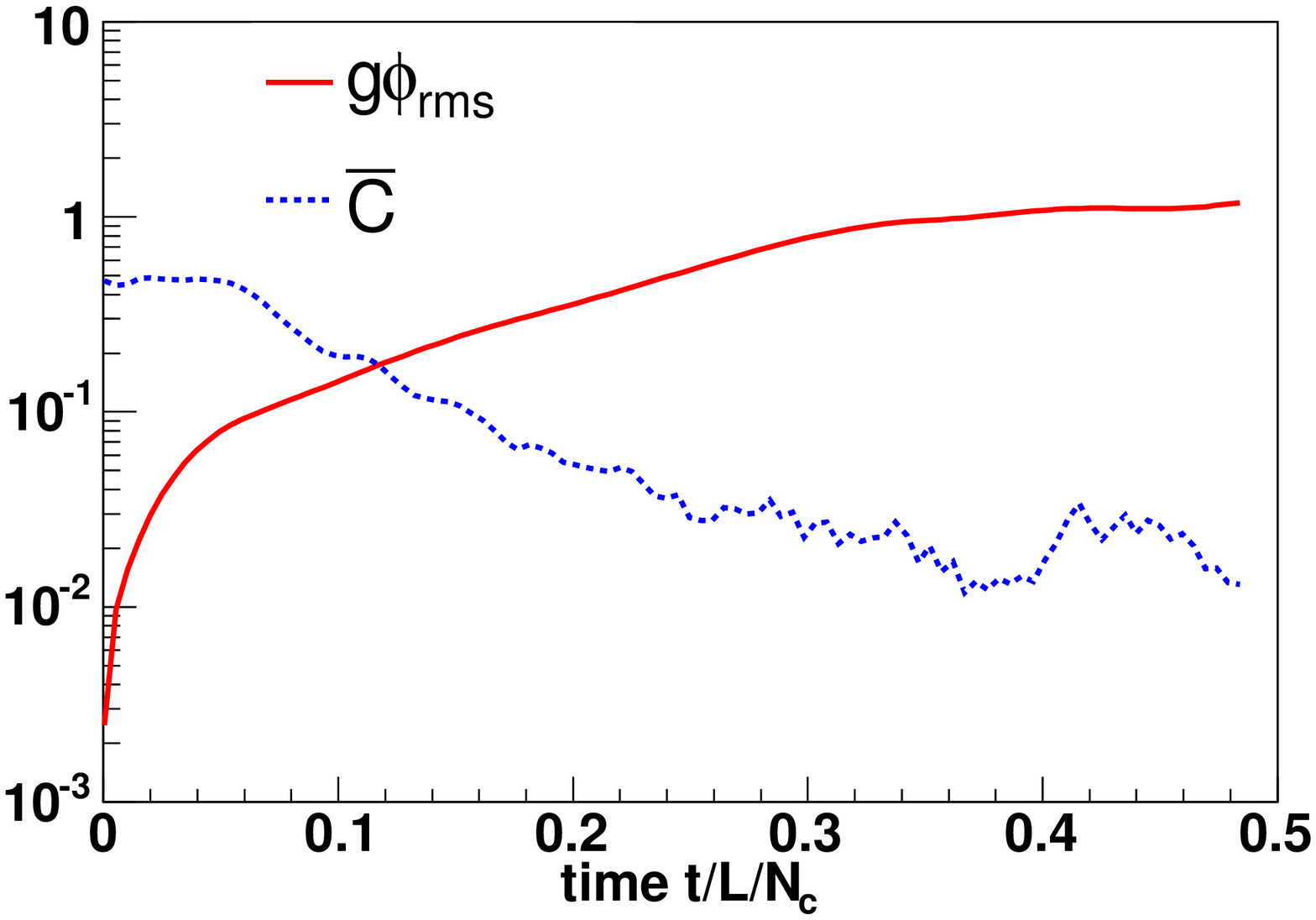}
\vspace{-0.4cm}
\caption{Temporal evolution of the functionals $\bar C$ and 
$\phi_{\rm rms}$ measured in GeV. The figure is taken from 
\cite{Dumitru:2005gp}.}
\label{fig-abel-dumitru}
\end{minipage}\hspace{2pc}%
\begin{minipage}{17pc}
\vspace{-2.0cm}
\includegraphics[width=20pc]{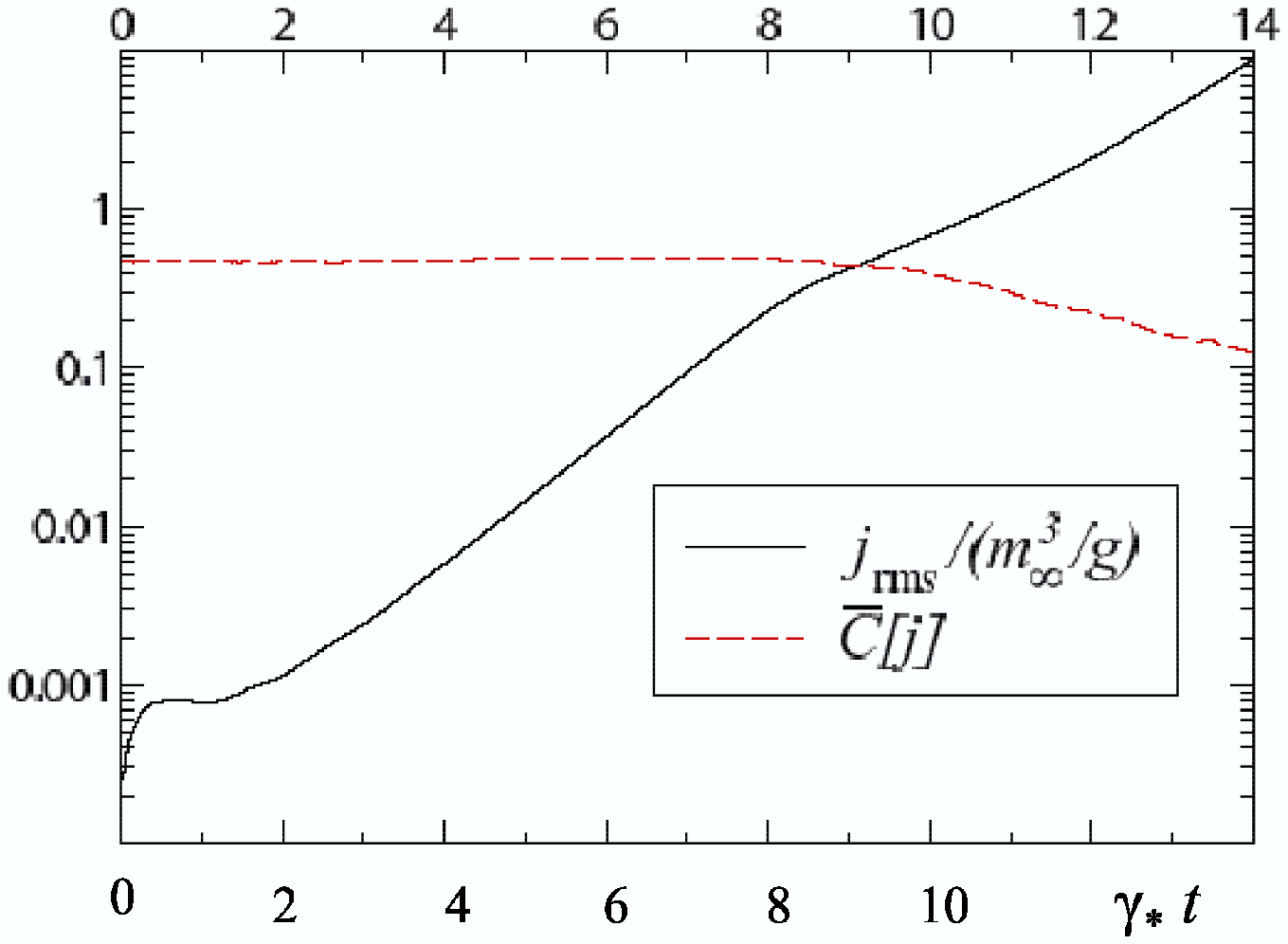}
\vspace{-4.3cm}
\caption{Temporal evolution of the (scaled) functionals $\bar C$ 
and $j_{\rm rms}$. The figure is taken from \cite{Rebhan:2004ur}.}
\label{fig-abel-rebhan11}
\end{minipage}
\end{figure}

In Fig.~\ref{fig-energy-rebhan11}, taken from \cite{Rebhan:2004ur}, 
the results of the Hard-Loop simulation performed in 1+1 dimensions 
are shown. One observes exponential growth of the energy density stored 
in fields and the energy density is dominated, as expected, by the 
magnetic field which is transverse to the direction of the momentum 
deficit. The growth rate of the energy density appears to be equal 
to the growth rate $\gamma^*$ of the fastest unstable mode. 
Fig.~\ref{fig-energy-dumitru}, which is taken from \cite{Dumitru:2005gp}, 
shows results of the classical simulation on the 1+1 dimensional lattice 
of physical size $L=40 \; {\rm fm}$. As in Fig.~\ref{fig-energy-rebhan11}, 
the amount of field energy, which is initially much smaller than the 
kinetic energy of all particles, grows exponentially and the magnetic 
contribution dominates. 

The Abelian (U(1)) and non-Abelian (SU(2)) results of the 1+1 
dimensional simulation presented in Fig.~\ref{fig-energy-dumitru} 
are remarkably similar to each other. The abelianization, explained 
in Sec.~\ref{sec-iso-abel}, appears to be very efficient in 1+1 
dimensions, as shown in 
Figs.~\ref{fig-abel-dumitru},~\ref{fig-abel-rebhan11}, taken from 
\cite{Dumitru:2005gp} and \cite{Rebhan:2004ur}, respectively.
The authors of \cite{Dumitru:2005gp} analysed the functionals
\ba
\label{C+phi}
\phi_{\rm rms} \equiv \sqrt{2\int_0^L \frac{dx}{L} 
{\rm Tr}[A^2_y + A^2_z]}
\;, \;\;\;\;\;\;
\bar C \equiv \int_0^L \frac{dx}{L} 
\frac{\sqrt{ {\rm Tr}[(i[A_y,A_z])^2]}}{{\rm Tr}[A_y^2+A_z^2]} \;,
\ea
which were introduced in \cite{Arnold:2004ih}. The quantities 
$j_{\rm rms}$ and $\bar C$, studied in \cite{Rebhan:2004ur} and 
shown in Fig.~\ref{fig-abel-rebhan11}, are fully analogous to 
$\phi_{\rm rms}$ and $\bar C$ defined by Eq.~(\ref{C+phi}) but 
the components of chromodynamic potential are replaced by the 
respective components of colour current. As seen in 
Figs.~\ref{fig-abel-dumitru},~\ref{fig-abel-rebhan11}, 
the field (current) commutator  decreases in time although the
magnitude of field (current), as quantified by $\phi_{\rm rms}$ 
($j_{\rm rms}$), grows.

It is worth mentioning that the functionals (\ref{C+phi}) defined
through the gauge potentials are gauge invariant provided the
potentials depend only of one time and one space variables and 
the gauge transformations preserve this property. Thus, the 
functionals (\ref{C+phi}) are well suited for $1+1$ dimensional 
simulations. However, the functionals (\ref{C+phi}) are {\em not} 
gauge invariant under general $1+3$ dimensional gauge 
transformations. When the potential components are replaced by 
the respective current components, as proposed in 
\cite{Rebhan:2004ur}, the functionals are gauge invariant not 
only under $1+1$ but also under $1+3$ dimensional transformations.

\begin{figure}
\begin{minipage}{16pc}
\includegraphics[width=16pc]{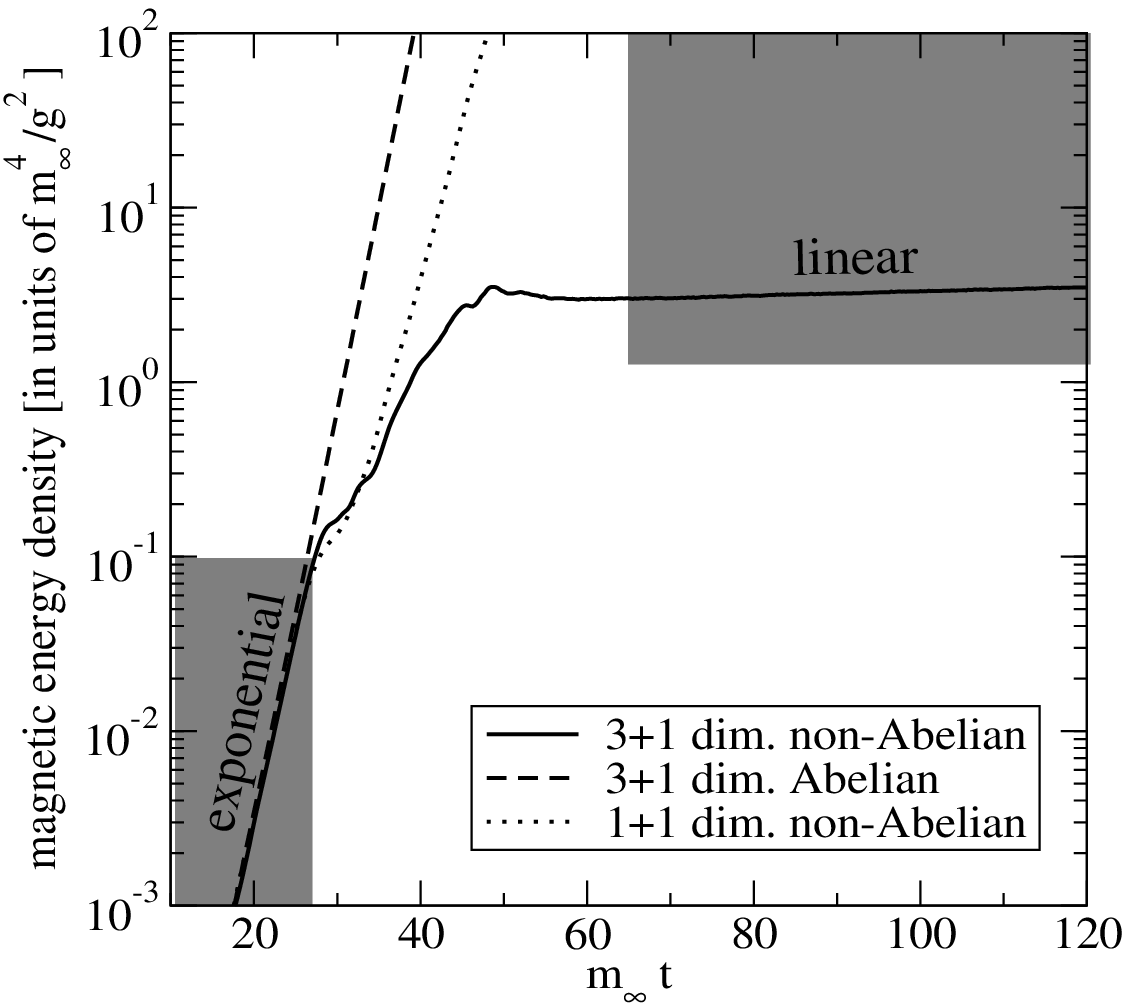}
\vspace{-0.9cm}
\caption{Time evolution of the (scaled) chromomagnetic energy 
density in the 1+3 dimensional simulation. The Abelian result 
and that of 1+1 dimensions are also shown. The figure 
is taken from \cite{Arnold:2005vb}.}
\label{fig-energy-arnold13}
\end{minipage}\hspace{2pc}%
\begin{minipage}{20pc}
\vspace{0.2cm}
\includegraphics[width=20pc]{fig4a.eps}
\vspace{-0.75cm}
\caption{Time evolution of the (scaled) energy density (split into 
various electric and magnetic components) of the chromodynamic 
field in the 1+1 and 1+3 simulations. `HL' denotes the total 
energy contributed by hard particles. The figure is taken 
from \cite{Rebhan:2005re}.}
\label{fig-energy-rebhan13}
\end{minipage}
\end{figure}

\begin{figure}
\begin{minipage}{17.5pc}
\includegraphics[width=17.5pc]{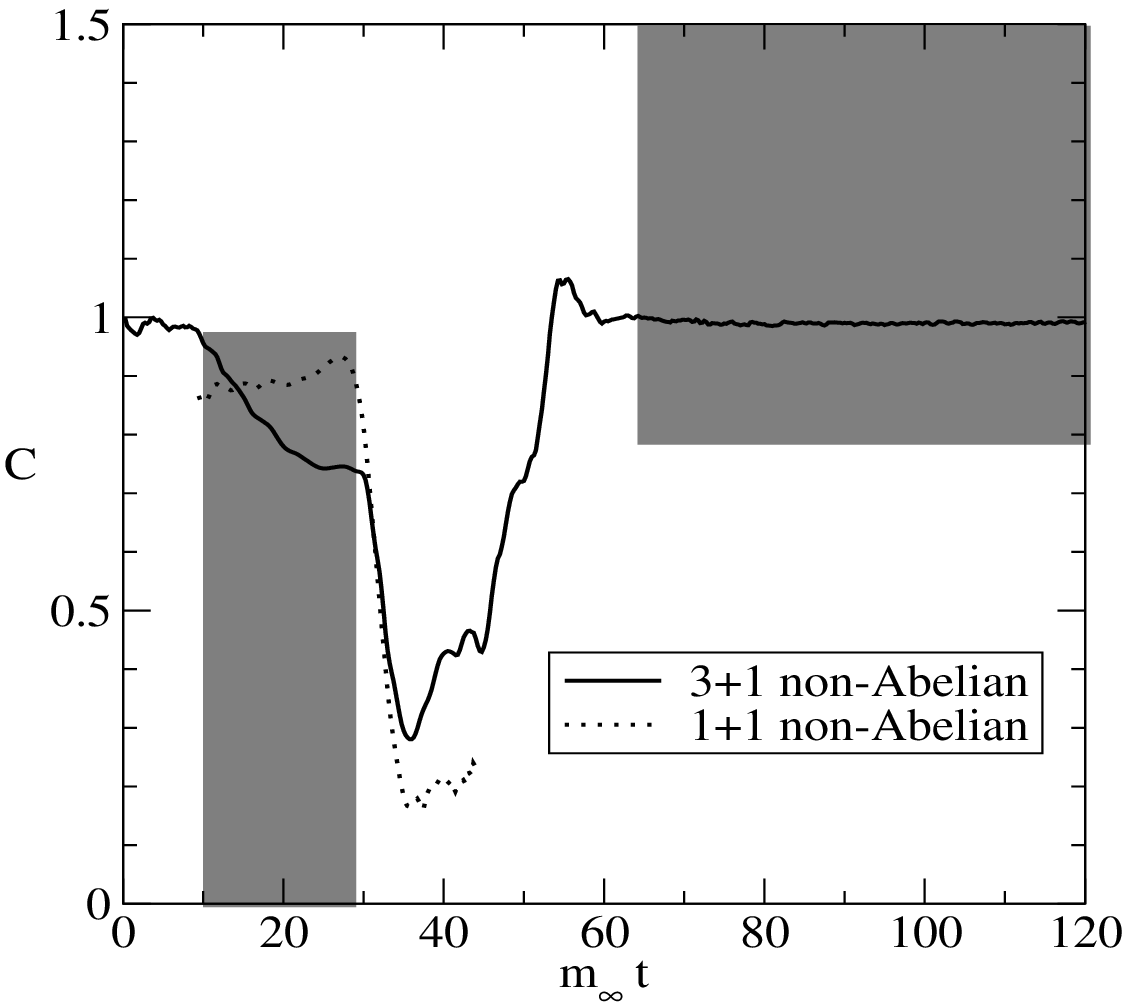}
\vspace{-0.9cm}
\caption{Temporal evolution of the field commutator quantified 
by $C$. The figure is taken from \cite{Arnold:2005vb}.}
\label{fig-abel-arnold13}
\end{minipage}\hspace{2pc}%
\begin{minipage}{18.5pc}
\vspace{-0.1cm}
\includegraphics[width=18.5pc]{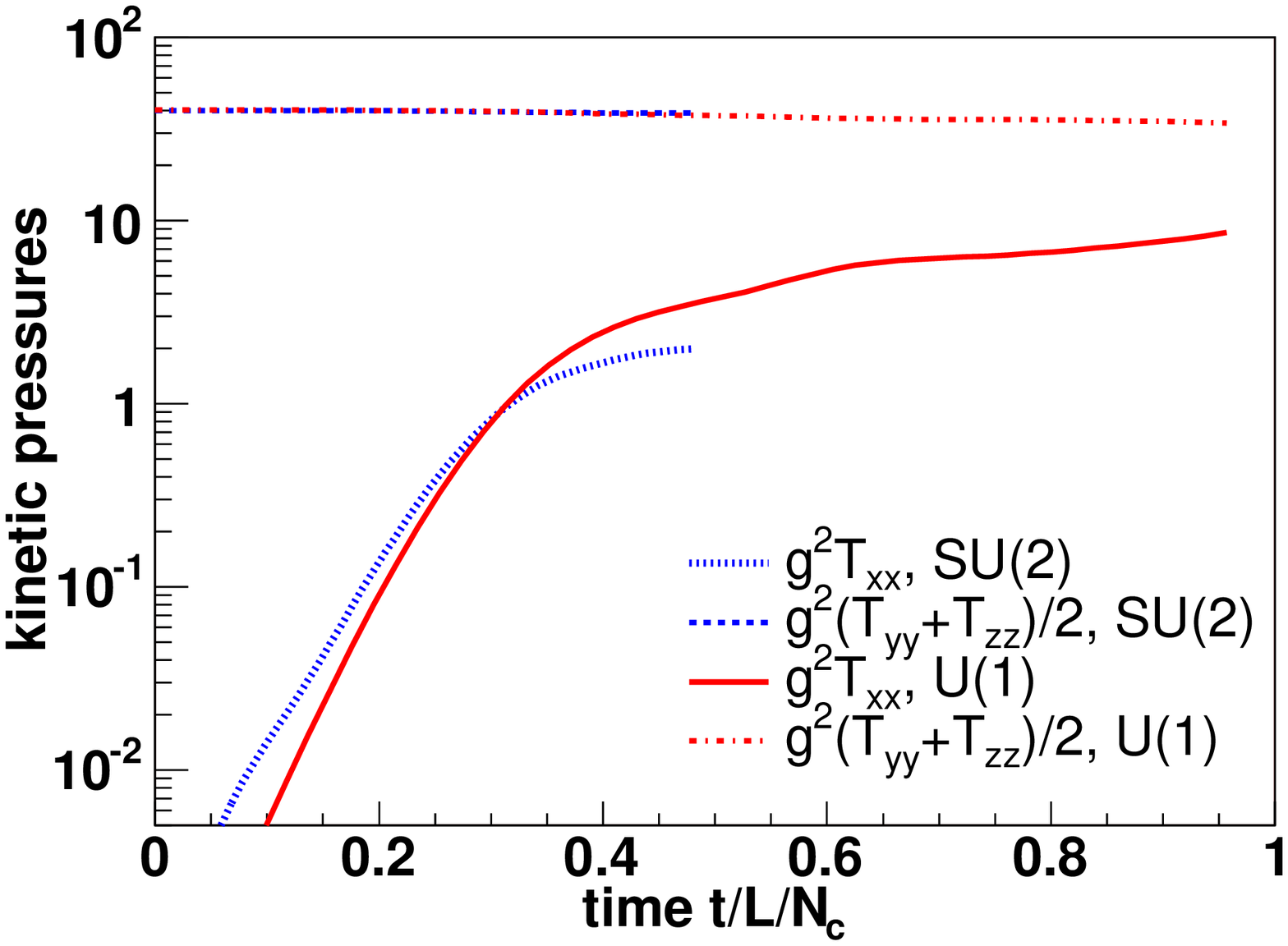}
\vspace{-0.3cm}
\caption{Temporal evolution of the energy-momentum tensor 
components $T^{xx}$ and $(T^{yy}+T^{zz})/2$. The Abelian 
and non-Abelian results are shown. The figure is taken from 
\cite{Dumitru:2005gp}.}
\label{fig-isotro}
\end{minipage}
\end{figure}

The results of the 1+3 dimensional simulations 
\cite{Arnold:2005vb,Rebhan:2005re} appear to be qualitatively 
different from those of 1+1 dimensions. As seen in 
Figs.~\ref{fig-energy-arnold13},~\ref{fig-energy-rebhan13}, 
taken from \cite{Arnold:2005vb} and \cite{Rebhan:2005re}, 
respectively, the growth of the field energy density is 
exponential only for some time, and then the growth becomes 
approximately linear. It appears that the regime changes when 
the field's amplitude is of order $k/g$ where $k$ is the 
characteristic field wave vector. Then, the non-Abelian effects 
start to be important. Indeed, Fig.~\ref{fig-abel-arnold13}, 
which is taken from \cite{Arnold:2005vb}, demonstrates that the 
abelianization is efficient in the 1+3 dimensional simulations 
\cite{Arnold:2005vb,Rebhan:2005re} only for a finite interval of
time. The commutator $C$ shown in Fig.~\ref{fig-abel-arnold13},
which is a natural generalization of the 1+1 dimensional commutator 
defined by Eq.~(\ref{C+phi}) with the current components instead
of the potential ones, first decreases but after some time
it starts to grow and returns to its initial value.

As discussed in two very recent papers \cite{Arnold:2005ef,Arnold:2005qs},
the physics of the late stage of instability development when 
the energy stored in the fields grows linearly with time is 
very interesting. Similarly to the turbulence, the unstable 
modes do not grow any more but due to non-Abelian interactions 
the energy provided by the particles is cascaded towards harder 
and harder modes. 

The effect of isotropization due to the action of the Lorentz 
force is nicely seen in the 1+1 dimensional classical simulation 
\cite{Dumitru:2005gp}. In Fig.~\ref{fig-isotro}, which is taken 
from \cite{Dumitru:2005gp}, there are shown diagonal components 
of the energy-momentum tensor
\ban
T^{\mu \nu} = \int \frac{d^3 p}{(2\pi)^3}\frac{p^\mu p^\nu}{E_p}
f({\bf p}) \;.
\ean
The initial momentum distribution is given by Eq.~(\ref{initial}),
and consequently $T^{xx}=0$ at $t=0$. As seen in Fig.~\ref{fig-isotro},
$T^{xx}$ exponentially grows. However, a full isotropy, which requires
$T^{xx}= (T^{yy} + T^{zz})/2$, is not achieved.

                                                                                
\section{Outlook and Final Remarks}
                                                                                

One wonders whether the presence of the instabilities at the early 
stage of relativistic heavy-ion collisions is experimentally observable. 
The accelerated equilibration is obviously very important though
it is only an indirect signal. It has been suggested 
\cite{Mike05,Muller:2005wi} that strong chromomagnetic fields 
generated by the instabilities can lead to a specific pattern 
of jet's deflections. This promising proposal, however, requires 
further studies.

Another idea has been formulated in \cite{Mrowczynski:2005gw}. 
The quark-gluon plasma, which is initially anisotropic,
is isotropized fast due to the magnetic instabilities. Such 
a non-equilibrium plasma manifests, as recently observed 
\cite{Arnold:2004ti}, an approximate hydrodynamic behaviour 
even before the equilibrium is reached. The point is that 
the structure of the ideal fluid energy-momentum 
tensor {\it i.e.} 
$T^{\mu \nu} = (\varepsilon + p) \, u^{\mu} u^{\nu} -p \, g^{\mu \nu}$, 
where $\varepsilon$, $p$ and $u^{\mu}$ is the energy density, 
pressure and hydrodynamic velocity, respectively, holds for an
arbitrary but isotropic momentum distribution. $\varepsilon$ 
and $p$ are then not the energy density and pressure but the 
moments of the distribution function which are equal the energy 
density and pressure in the equilibrium limit. Since the tensor 
$T^{\mu \nu}$ always obeys the continuity equation 
$\partial_\mu T^{\mu \nu} =0$, one gets an analogue of the Euler
equation. However, due to the lack of thermodynamic equilibrium
there is no entropy conservation and the equation of state is
missing.

The azimuthal fluctuations have been argued \cite{Mrowczynski:2005gw} 
to distinguish the approximate hydrodynamics -- characteristic 
for the instabilities driven isotropization -- from the real 
hydrodynamics describing a system which is in a local thermodynamic 
equilibrium, as advocated by proponents of the strongly coupled plasma 
\cite{Shuryak:2004kh}. Non-equilibrium fluctuations are usually 
significantly larger than the equilibrium fluctuations of the same 
quantity. A specific example of such a situation is given in 
Sec.~\ref{sec-seeds} where the current fluctuations in the anisotropic 
system are discussed. Thus, one expects that the (computable) 
fluctuations of $v_2$ produced in the course of real hydrodynamic 
evolution are significantly smaller than those generated in the 
non-equilibrium quark-gluon plasma which is merely isotropic. It 
should be stressed here that the elliptic flow is generated in 
the collision relatively early stage when there is a large 
configuration-space asymmetry of the colliding system. Since 
a measurement of $v_2$ fluctuations is rather difficult, it was also 
argued \cite{Mrowczynski:2005gw} that an integral measurement of 
the azimuthal fluctuations can help as well to distinguish the 
equilibrium from non-equilibrium fluctuations.  

Further suggestions of detectable signals of the instabilities 
are very much needed. However, an experimental verification will 
certainly require much better theoretical understanding of the 
equilibration process. Although an impressive progress has been 
recently achieved, the numerical simulations 
\cite{Rebhan:2004ur,Dumitru:2005gp,Arnold:2005vb,Rebhan:2005re}
are still quite far from the situation which occurs in relativistic 
heavy-ion collisions. A realistic simulation should properly 
take into account an initial state of colliding nuclei; the 
system's expansion, which slows down or even cuts off the 
instabilities growth \cite{Randrup:2003cw,Arnold:2003rq}, needs 
to be included; the simulation should be $1+3$ dimensional as the 
dimensionality really matters \cite{Arnold:2005vb,Rebhan:2005re}; 
the effect of back reaction of fields on the particles has to 
be taken into account to observe the system's isotropization. 

Some of the above requirements are met by a very recent numerical 
study \cite{Romatschke:2005pm,Romatschke:2005ag} where the initial
state corresponds to the Colour Glass Condensate \cite{Iancu:2003xm} 
where small $x$ partons of large occupation numbers, which dominate 
the wave functions of incoming nuclei, are treated as classical 
Yang-Mills fields. As already mentioned, hard modes of the classical 
fields play the role of particles here. The study shows that the 
instabilities, identified as the Weibel modes, are indeed generated 
when the system of Yang-Mills fields representing colliding nuclei 
expands into the vacuum. However, the unstable mode growth is, as 
argued in \cite{Randrup:2003cw,Arnold:2003rq}, slowed down.

Understanding of the late stage of the instability growth, 
when fields are of large magnitude, is a real theoretical
challenge. The mechanism of instability saturation is not well
known even in the electron-ion plasma, see {\it e.g.} a
recent paper \cite{Kato:2005wv}. Non-linear effects, 
in particular those of non-Abelian nature, are then essential.
Except for the classical simulations 
\cite{Dumitru:2005gp,Romatschke:2005pm,Romatschke:2005ag}, 
the evolution of anisotropic quark-gluon plasma has been 
studied within the Hard Loop approximation. An attempt to
go beyond it has been undertaken in \cite{Manuel:2005mx}
where the higher order terms of the effective potential 
of the anisotropic system have been computed. Since these 
terms can be negative, the instability is then driven not 
only by the negative quadratic term but by the higher order 
terms as well. However, before a real progress in the strong 
field domain can be achieved, one still needs a better insight 
into the Hard Loop dynamics which has appeared to be very rich 
\cite{Arnold:2005vb,Rebhan:2005re,Arnold:2005ef}. 

In summary, the magnetic instabilities provide a plausible 
explanation of the surprisingly short equilibration time 
observed in relativistic heavy-ion collisions. The explanation
does not require a strong coupling of the quark-gluon plasma.
Fast isotropization of the system is a distinctive feature of 
the instabilities driven equilibration. Two signals of the 
instabilities have been suggested but quantitative predictions 
are lacking. New ideas are certainly needed. In spite of the 
impressive progress, which has been achieved recently, a theoretical 
description of the unstable quark-gluon plasma requires further 
improvements.

\begin{acknowledgements}

I am indebted to Adrian Dumitru, Cristina Manuel, Toni Rebhan, 
Paul Romatschke, Mike Strickland, Raju Venugopalan, Stephen Wong 
and Larry Yaffe for comments on the manuscript. A support by the 
Virtual Institute VI-146 of Helmholtz Gemeinschaft is also 
gratefully acknowledged.

\end{acknowledgements}
%
%

\end{document}